\documentclass[12pt]{IOPART}
\usepackage{graphicx}
\begin{document}

\title[Fr\"{o}hlich superconductivity in high magnetic fields]{Experimental
evidence for Fr\"{o}hlich superconductivity in high magnetic fields}

\author{N.~Harrison$^1$,\footnote[3]{To
whom correspondence should be addressed (nharrison@lanl.gov)}
C.~H.~Mielke$^1$, J.~Singleton$^1$,
J.S.~Brooks$^2$ and M.~Tokumoto$^3$}

\address{$^1$National High Magnetic Field Laboratory, LANL, MS-E536, Los
Alamos, New Mexico 87545, USA}

\address{$^2$National High Magnetic Field Laboratory, Florida State
University, Tallahassee, Florida 32310, USA}

\address{$^3$Electrotechnical Laboratory, Tsukuba, Ibaraki 305, Japan}

\begin{abstract}
Resistivity and irreversible magnetisation data taken within
the high magnetic field CDW$_x$ phase of
the quasi-two-dimensional organic metal
$\alpha$-(BEDT-TTF)$_2$KHg(SCN)$_4$
are shown to be consistent with a field-induced inhomogeneous
superconducting phase.
In-plane skin depth measurements show that the resistive
transition on entering the CDW$_x$ phase is both isotropic and
representative of the bulk.
\end{abstract}

\submitto{\JPC}
\maketitle
Prior to the development of the BCS theory of superconductivity
\cite{BCS}, Fr\"{o}hlich proposed a novel form of superconductivity in
one-dimensional (1D) metals due to spontaneously sliding
charge-density waves (CDWs)~\cite{frohlich}. CDWs were subsequently
discovered in many materials containing quasi-one-dimensional
(Q1D) Fermi-surface sections~\cite{gruner}; in such systems,
a gap opens up over part or all of the Q1D Fermi-surface section(s)
due to a nesting instability~\cite{gruner}. However, 
this gap, and the pinning of
the CDW to impurities, generally prevent a CDW from contributing to the
electrical conductivity, so that the electrical transport properties
are dominated by any remaining normal carriers~\cite{gruner}.
Conduction by sliding CDWs due to their depinning may occur
under large electric fields, but is accompanied by considerable
dissipation~\cite{gruner}. In this respect, the organic conductor
$\alpha$-(BEDT-TTF)$_2$KHg(SCN)$_4$ (together with the analogue
compounds in which K has been substituted by Tl or Rb) may be an
exception \cite{oshima}. This highly anisotropic layered metal
exhibits what is thought to be a CDW with an exceptionally low
transition temperature $T_{\rm p}\sim$~8~K
\cite{sasaki,miyagawa,mckenzie,biskup,harrison1}. At applied magnetic
fields ($\mu_0H$) greater than 23~T and temperatures ($T$) less than
2~K, $\alpha$-(BEDT-TTF)$_2$KHg(SCN)$_4$ possesses a phase which
carries dissipationless currents \cite{harrison2,harrison3}.

In this Letter, we present compelling evidence that these currents are
conveyed by the CDW itself. Measurements of the skin depth reveal that
the currents propogate deep into the bulk of the material.
Furthermore, the drop in resistivity with decreasing temperature is
shown to be isotropic, possessing a form characteristic of a
transition into an inhomogeneous superconducting phase.

Fr\"{o}hlich's original model
neglects interactions with impurities \cite{frohlich}; these
spatially pin the CDW, preventing the
spontaneous sliding motion required for superconductivity
\cite{gruner}. This pinning is often conveniently represented by a
``particle'' in a sinusoidally-varying `washboard' potential
\cite{gruner}. There are, however, limits in which this model should
fail. For example, if the ``particles's'' zero point energy
$\varepsilon_0=\hbar^2Q^2/4m^\ast$ exceeds the potential depth $V$,
the washboard potential can no longer confine the CDW. Here, $Q$ is
the wavevector (or nesting vector) representing the spatial periodicity of the
potential and $2m^\ast$ is the paired electron-hole mass, enhanced due to
interactions between the CDW and the crystal lattice \cite{gruner}.
The degree to which $m^\ast$ is renormalised with respect to the
effective mass characteristic of the unperturbed band ($m_{\rm b}$)
is proportional to the square of the gap that opens at the Fermi
energy on formation of the CDW \cite{gruner}.

\begin{figure}[htbp]
\centering
\includegraphics[height=12cm]{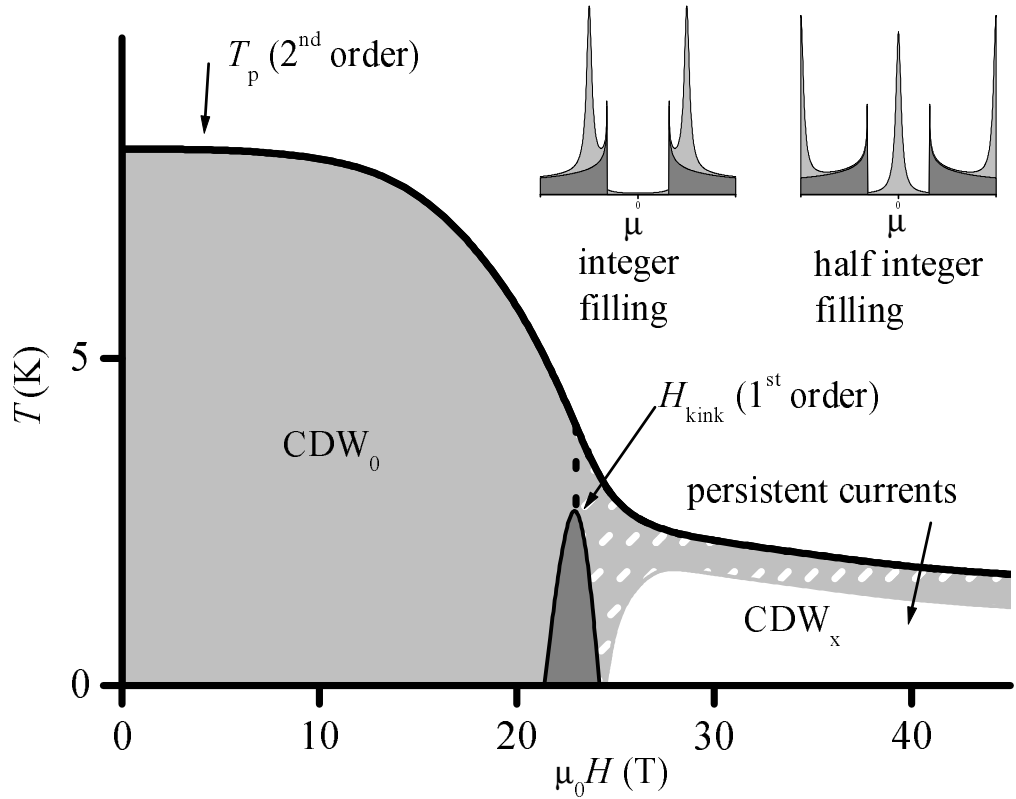}
\caption{Notional phase diagram of $\alpha$-(BEDT-TTF)$_2$KHg(SCN)$_4$
constructed from theoretical models (References [7] and [12]) and data
accumulated in this and other works (References [10,13,14,16])). The solid
line
represents a second order transition into the CDW phase (light shading) with
a dotted line [together with the region of hysteresis (heavy shading)]
representing a first order transition between the proposed CDW$_0$
(solid shading) and CDW$_x$ (hatched shading) phases. Persistent
currents (white region) are observed only within the CDW$_x$
phase. In the top right-hand corner, we show the density of states
resulting from CDW formation within the 1D bands (heavy shading) together
with the contributions from the Landau levels of the 2D band (light
shading), both at integer and half integer filling factors. $\mu$
represents the position of the chemical potential.
}
\label{diagram}
\end{figure}

The energy gap in $\alpha$-(BEDT-TTF)$_2$KHg(SCN)$_4$ at
$T=\mu_0H=$~0 can be estimated using the BCS relation
$2\Delta_0=3.52 k_{\rm B}T_{\rm p}\approx$~3~meV
\cite{frohlich,gruner,harrison2}. It has been suggested that this gap
is closed by the Zeeman energy when $\mu_0\mu_{\rm
B}H/\sqrt{2}\approx 2\Delta_0$, corresponding to $\mu_0H\approx$~23~T
\cite{mckenzie,harrison1}; this causes a first-order phase transition
(the ``kink'' transition) into a proposed spatially modulated CDW$_x$
phase \cite{mckenzie,harrison1,harrison2,zanchi,christ,qualls}, (see
notional phase diagram in Fig. \ref{diagram}).
$\alpha$-(BEDT-TTF)$_2$KHg(SCN)$_4$ is therefore unusual in two
respects: (1) it exhibits the only known CDW state in which
$\Delta_0$ is sufficiently small for such a transition to have been shown to
occur in
experimentally-accessible magnetic fields \cite{almeida}; and (2) the
reduced gap
$2\Delta_x\approx$~1~meV that characterizes the CDW$_x$ phase
\cite{harrison2} is two orders of magnitude smaller than those in
typical CDW materials \cite{gruner}. The smallness of the gap leads to
a negligibly small enhancement of
$m^*$ with respect to $m_{\rm b}$ (in contrast to that in more typical
CDWs), leading to estimates of $\varepsilon_0$ in the range
$10^2$~-~$10^3$ meV (using $Q\approx$~7~$\times$~10$^{9}$~m$^{-1}$
from Reference \cite{singleton}). Hence $\varepsilon_0$ greatly
exceeds both $2\Delta_x$ and $V$, given that $V$ must be less than
$\Delta_x$ for the CDW to be stable \cite{gruner}. These arguments
lead one to expect that the CDW$_x$ phase may undergo the
spontaneous sliding necessary for Fr\"{o}hlich superconductivity
\cite{frohlich}. In the following, we will show that the resistivity
and the magnetisation of $\alpha$-(BEDT-TTF)$_2$KHg(SCN)$_4$ within
the CDW$_x$ phase do indeed exhibit behaviour characteristic of an
inhomogeneous superconductor.

\begin{figure}[htbp]
\centering
\includegraphics[height=16cm]{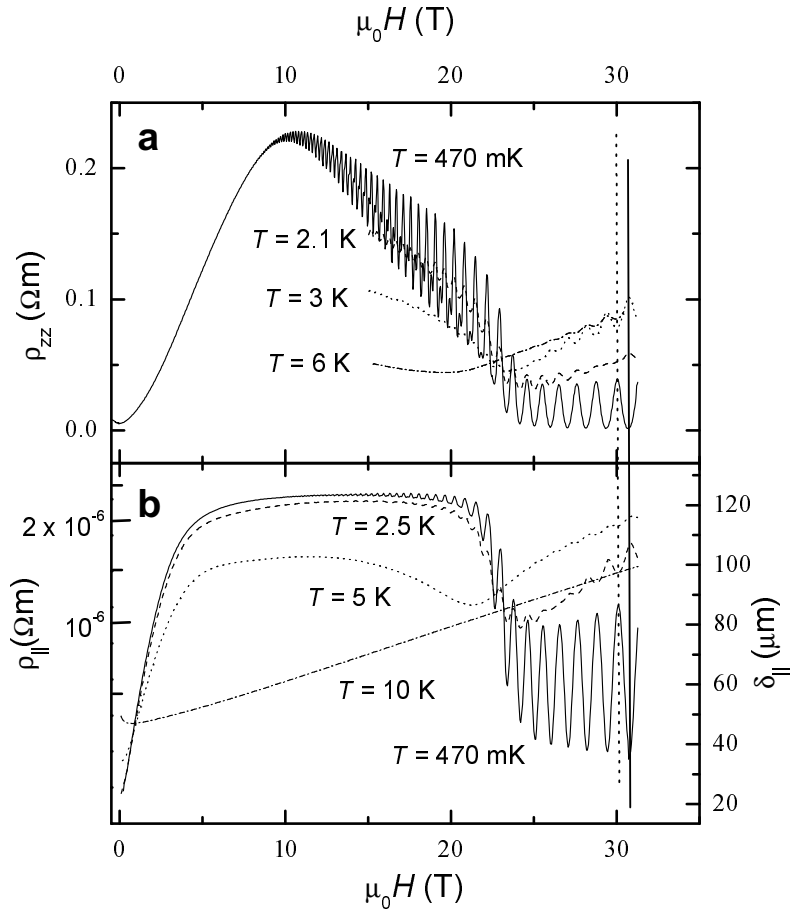}
\vspace{-1.0in}
\caption{Examples of the interplane and in-plane resistivity
components, $\rho_{zz}$ and $\rho_{||}$, of
$\alpha$-(BEDT-TTF)$_2$KHg(SCN)$_4$
plotted versus dc magnetic field at different temperatures.
$\rho_{zz}$ (a) is measured using the conventional four wire transport
technique while $\rho_\|$ (b) is estimated from the skin depth
$\delta_\|$ (note that the $\rho_\|$ axis is non-linear). The angle
between the interplane $b$-axis of the sample and the dc magnetic
field is $\sim$~7$^\circ$. The solid vertical line indicates an example
of a magnetic field value where $\nu$ is an integer while the dotted line
indicates one where $\nu$ is a half-integer.
}
\label{rho}
\end{figure}

Figure \ref{rho} compares different measurements of the resistivity
of $\alpha$-(BEDT-TTF)$_2$KHg(SCN)$_4$; the interplane resistivity
component $\rho_{zz}$ is measured using conventional four-terminal
transport \cite{harrison2}. However, the highly anisotropic
conductivity of $\alpha$-(BEDT-TTF)$_2$KHg(SCN)$_4$ makes it very
difficult to measure the average in-plane resistivity $\rho_\|$ using
such techniques \cite{singleton}. By contrast, the radio-frequency
skin depth $\delta_\|=\sqrt{2\rho_\|/\mu_0\omega}$ enables both
$\rho_\|$ and the extent to which the currents penetrate the sample
to be determined simultaneously \cite{coffey}. The single-crystal
sample is placed inside the inductor of a tunnel diode oscillator
(TDO) tank circuit \cite{coffey}, with the oscillating magnetic
field $\tilde{\bf h}$ perpendicular to the conducting layers.
Screening within the sample results in a reduction in inductance, and
consequently to an increase in the resonant frequency $\omega/2\pi$.
Changes in frequency can be measured to high precision, with
$\delta_\|$ being approximately related to $\omega$ by
$\delta_\|\approx [\frac{1}{2}+(\omega-\omega_0)/\eta\omega_0]r$.
Here, $\omega_0/2\pi=$~38.975~$\pm$~0.005~MHz is the resonant
frequency without the sample, $\eta=$~0.053~$\pm$~0.002 is the
effective filling factor of the sample with respect to the total
inductance of the circuit and $r=$~250~$\pm$~50~$\mu$m is the
effective sample radius \cite{coffey}.

Figure \ref{rho}b shows $\rho_\|$ extracted from $\delta_\|$ at
several different temperatures. At temperatures $T>$~8~K,
$\alpha$-(BEDT-TTF)$_2$KHg(SCN)$_4$ displays the positive
magnetoresistance exhibited by most organic metals \cite{singleton}.
At temperatures below 8~K, the magnetoresistance develops
a semiconducting-type behaviour for 3~$<\mu_0H<$~23~T ({\it i.e.} $\rho_\|$
increases with decreasing $T$), enabling $\tilde{\bf h}$ to fully
penetrate; however, above 23~T, the resistivity is lower, the change
becoming more distinct as the temperature decreases. The
Shubnikov-de~Haas (SdH) oscillations in Fig. \ref{rho} are due to an
additional two-dimensional (2D) band \cite{mori} that, while remaining
largely unaffected by the CDW order, becomes quantized in a series of
sharply-defined Landau levels at high magnetic fields
\cite{shoenberg}. Since the number of filled levels $\nu$ is
proportional to $1/B$ (owing to an increase in the degeneracy of each
level with magnetic induction $B\approx\mu_0H$) \cite{shoenberg}, the
field can be used to tune the position of the chemical potential
relative to the Landau levels; at integer filling factors $\nu=F/B$,
where $F$ is the frequency of the SdH oscillations, the chemical
potential resides in a Landau gap between the highest filled and
lowest empty Landau level.

The data in Fig. \ref{rho} are very important, because they show that
in almost all respects, including the detailed phase behaviour of the
SdH oscillations, the variation of $\rho_\|$ (estimated from
$\delta_\|$) with $\mu_0H$ and $T$ at high fields is identical to that
of $\rho_{zz}$. (Of course $\rho_{zz}$ is several orders of magnitude
greater than $\rho_\|$ owing to the anisotropy of the band structure
\cite{mori}). We can therefore assert that the changes taking place in
the resistivity as a function of $\mu_0H$ and $T$ are intrinsic, bulk
properties of $\alpha$-(BEDT-TTF)$_2$KHg(SCN)$_4$.

\begin{figure}[htbp]
\centering
\includegraphics[height=18cm]{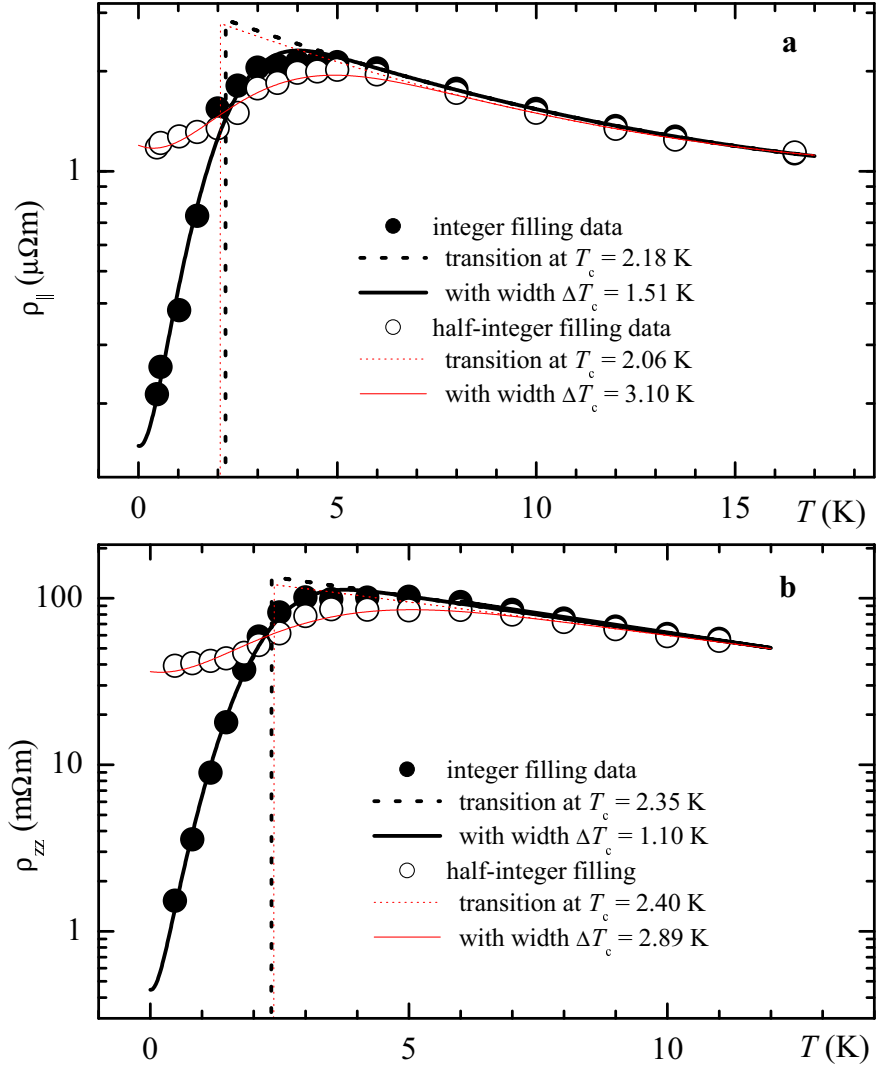}
\vspace{-0.5in}
\caption{Temperature dependence of the resistivity. (a), The in-plane
resistivity $\rho_\|$ (filled circles) extracted from $\delta_\|$ at
integer filling factors (at $\mu_0H \approx 30$~T), together with a
resistivity extrapolation (using a polynomial fit) from within the
normal phase down to the midpoint of the transition $\bar{T}_{\rm c}$
(heavy dotted line). Below $\bar{T}_{\rm c}$ the resistivity is zero.
On multiplying the extrapolated resistivity by the function
$\frac{1}{2}({\rm erf}[(T-\bar{T}_{\rm c})/\Delta T_{\rm c}]
+{\rm erf}[(-T-\bar{T}_{\rm c})/\Delta T_{\rm c}]+2)$, we obtain excellent
agreement (heavy solid line) with the experimental data (filled
circles). The same procedure is also performed on the data at
half-integer filling factors (but with open symbols and thinner
lines), leading to a very similar
value of $\bar{T}_{\rm c}$ but broader transition.
(b), The same analysis repeated on
the interplane resistivity data taken from Reference [10].
}
\label{Tdep}
\end{figure}

The temperature dependences of $\rho_{zz}$ and $\rho_\|$ are shown in
Fig. \ref{Tdep} for both integer and half-integer $\nu$. Remarkably,
the temperature dependence of $\rho_{zz}$ and $\rho_\|$ at integer
$\nu$ is exactly the same as a transition into an inhomogeneous
superconducting phase \cite{testardi,maza,veira}. In such systems,
the (normally sharp) transition into the zero resistivity ground state
becomes statistically broadened by a gaussian distribution with
standard deviation $\Delta T_{\rm c}$ about a mean critical
temperature $\bar{T}_{\rm c}$ \cite{testardi,maza,veira}. As can be
seen in Fig. \ref{Tdep}, the convolution of a sharp transition into a
zero resistivity state with gaussian broadening 
({\it i.e.} the error function) provides an excellent
fit to both the $\rho_{zz}$ and $\rho_\|$ data at integer filling
factors. This implies that an increasing fraction of the sample no
longer contributes to the resistivity as $T$ is lowered, in a manner
consistent with the carriers (within this fraction) having condensed
into a superconducting phase. As $\Delta T_{\rm c}\approx$~1~K is a
significant
fraction of $\bar{T}_{\rm c}\approx$~2~K, the transition is never entirely
complete, so that the total resistivity of the inhomogeneous phase
does not actually fall to zero at $T=$~0. Instead, the resistivity
drops off exponentially in the limit $T\rightarrow$~0 ({\it i.e.}
$\rho\propto{\rm e}^{T/T_0}$) in a manner similar to that observed in
several well-known examples of inhomogeneous superconductors
\cite{hsu,bednorz}. This resistivity model can also be fitted to the
data at half-integer filling factors with a very similar
$\bar{T}_{\rm c}$, but with the
transition width $\Delta T_{\rm c}$ being considerably broadened.
The fact that the midpoint of the transition $\bar{T}_{\rm c}$
is virtually unaffected by the Landau-level
filling factor of the 2D quasiparticles
(see Figure~1 inset) is consistent with the gap formation
being solely determined by the Q1D sheets.
The Q2D Landau density of states appears only to modulate the
width of the transition, possibly as a result
of its influence on the scattering rate;
there is an increased density of available states at
the chemical potential
at half-integer filling factors(see Fig. \ref{diagram}
inset)~\cite{singleton}.

\begin{figure}[htbp]
\centering
\includegraphics[height=12cm]{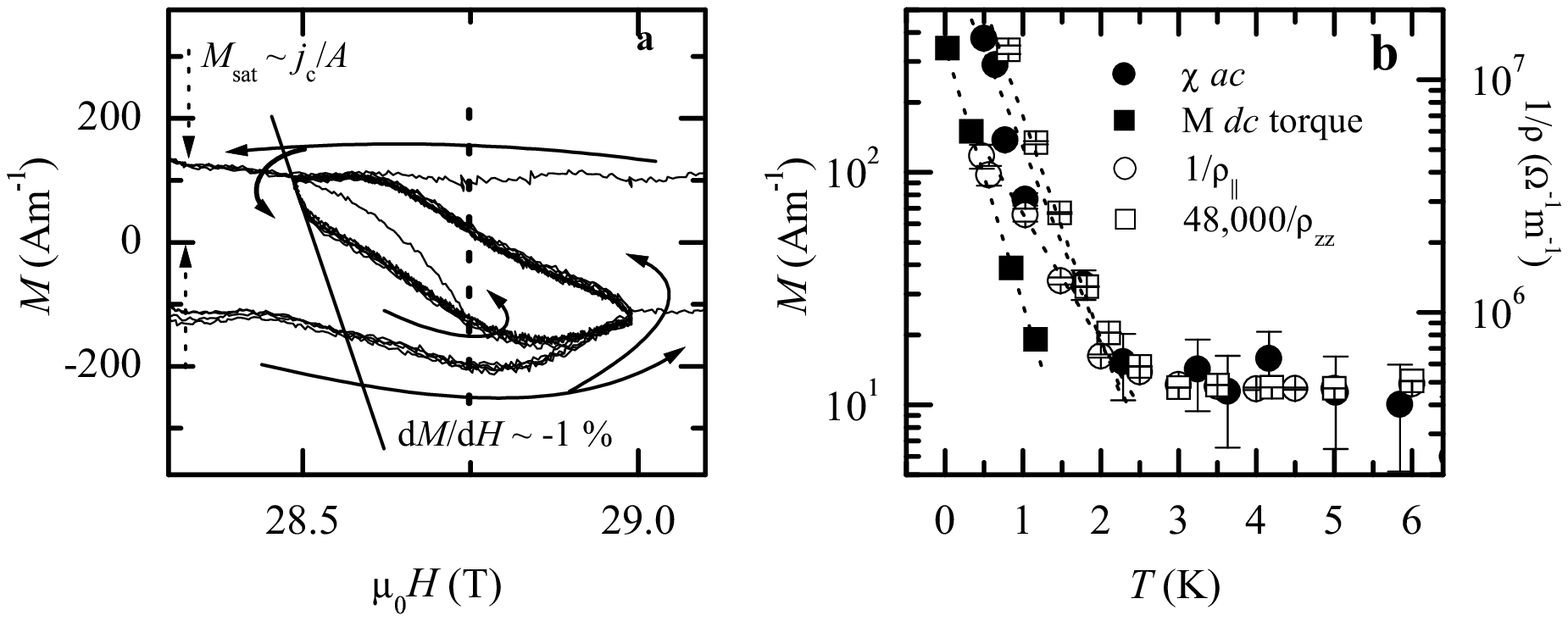}
\vspace{-1.5in}
\caption{Examples of magnetic measurements made on
$\alpha$-(BEDT-TTF)$_2$KHg(SCN)$_4$. (a), An example of a hysteresis
loop in the irreversible magnetisation $M$ estimated from the
magnetic torque measured at an angle of $\sim$~7$^\circ$ between the
inteplane $b$-axis of the sample and the magnetic field. Solid arrows
indicate the direction of sweep of the magnetic field, often swept
over the same interval in field to show reproducibility. The dashed
arrows indicate the saturated irreversible magnetisation $M_{\rm
sat}$, while the inclination of the solid straight line indicates the
negative differential susceptibility $\partial M/\partial H$. The
vertical dotted line indicates the field at which $\nu$ is half
integer. (b), Comparisons of the irreversible magnetisation of the
currents measured by means of ac susceptibility and magnetic torque
(filled symbols) with the conductivity $1/\rho$ (open symbols),
versus $T$, with dotted lines drawn merely to guide the eye. In an
inhomogeneous superconductor, both approximately scale with the
fraction of the sample in which currents are conveyed without
dissipation.
}
\label{mag}
\end{figure}

Finally, we turn to the magnetisation, which exhibits a behaviour
entirely consistent with that of an inhomogeneous superconductor in
the CDW$_x$ phase. Figure \ref{mag}a shows an example of a hysteresis
loop in the irreversible magnetisation $M$ obtained using
a torque magnetometer (see Reference \cite{harrison2}
for a detailed description of this technique); two important features of
this loop are consistent with Bean's critical state model for vortex
pinning in type II superconductors \cite{bean}. Firstly, the
irreversible magnetic susceptibility $\partial M/\partial H$ is
diamagnetic ({\it i.e.} negative) or reversing the direction of sweep
of $H$ \cite{harrison2}; this effect can only be ascribed to
circulating currents
\cite{harrison2,harrison3} (while circulating currents are not
expected in a purely 1D CDW system, they can occur in 2D CDW systems such as
$\alpha$-(BEDT-TTF)$_2$KHg(SCN)$_4$ \cite{singleton,bonkers}).
Secondly, the magnitude of the hysteresis
saturates at a critical value $M_{\rm sat}$ enabling us to identify a
critical current density $j_{\rm c}$ \cite{harrison2,bean}. Because
the hysteresis occurs at all filling factors $\nu$, the currents
cannot originate from the quantum Hall effect (QHE), as earlier
experiments had appeared to suggest \cite{singleton}. In the limit
$\nu\gg$~1 that applies here, the QHE is expected to operate only at
integer filling factors \cite{chakraborty}.

Thus, our reasons for attributing these effects within the high
magnetic field phase of $\alpha$-(BEDT-TTF)$_2$KHg(SCN)$_4$ to
Fr\"{o}hlich superconductivity \cite{frohlich} can be summarized as
follows: (1) the semiclassical model for CDW pinning is expected to
fail \cite{gruner}; (2) both the magnetisation and the resistivity
display effects found only in superconductors; the induced currents
are consistent with Bean's critical state model
\cite{harrison2,bean}, while the resistivity has exactly the form of a
transition into an inhomogeneous superconducting phase
\cite{testardi,maza,veira}; (3) these effects are observed at {\it
all} filling factors for $\mu_0H>$~23~T; and (4) only a single
resistive transition is observed as a function of $T$ on entering the
region in $H$ and $T$ over which the currents are observed, with the
midpoint $\bar{T}_{\rm c}$ of this transition (being virtually
independent of the 2D density of states) coinciding with
the transition into the CDW$_x$ phase predicted by theory
\cite{mckenzie,zanchi}.

The small diamagnetic fraction ($-\partial M/\partial H\sim$~1~\%)
\cite{harrison2}, large skin depth ($\delta_\|>$~30~$\mu$m, being an
appreciable fraction of $r$) and broadness of the resistive
transition ($\Delta T_{\rm c}\sim 1-3$~K) all suggest that the
persistent currents are confined to many small islands or filaments
throughout the sample surrounded by regions that are normal. Fig.
\ref{mag}b provides further support for this picture, where it is
apparent that both the size of the irreversible magnetisation $M$ and
the conductivity $1/\rho$ obey an approximate ${\rm e}^{-T/T_0}$ law
in the limit $T\rightarrow$~0 (see Fig. \ref{mag}b). It is interesting
that the transition width $\Delta T_{\rm c}$ is so broad for a
material that should most definitely be in the clean limit. We can
infer this from the fact that the mean free path $l=v_{\rm
F}\tau\sim$~4000~\AA ~(estimated from the results of de Haas-van Alphen
measurements \cite{harrison4}, where $v_{\rm F}\sim$~80,000~ms$^{-1}$
is the Fermi velocity) greatly exceeds the coherence length
$\zeta_0=\hbar v_{\rm F}/\pi\Delta\sim$~200~\AA (conservatively
estimated using the BCS formula \cite{BCS,gruner}). Should
$\alpha$-(BEDT-TTF)$_2$KHg(SCN)$_4$ possess a vortex state,
inhomogeneity could be a consequence of the transition being
broadened by a magnetic field in a similar manner to that observed in
other highly anisotropic superconductors \cite{tinkham,ishiguro}.
Another possibility to consider is that, unlike Cooper pairs in
conventional superconductors \cite{BCS}, a CDW charge modulation
experiences additional Coulomb interactions with impurities
\cite{gruner}. While the Coulomb force may not necessarily be able to
pin the CDW condensate if $\Delta$ is too small, it may still affect
the stability of the CDW order should $V$ become comparable to
$\Delta$ (see Reference \cite{gruner}).

In conclusion, both transport and magnetisation measurements
within the low temperature, high magnetic field CDW$_x$ phase of
$\alpha$-(BEDT-TTF)$_2$KHg(SCN)$_4$ are shown to be consistent with
inhomogeneous superconductivity, realised in small
filaments or islands throughout the bulk of the sample. The
locality of this effect with respect to the CDW phase diagram in $H$ and $T$
leads us to propose that Fr\"{o}hlich superconductivity (or
perhaps
a 2D variant thereof \cite{bonkers}) is realised in this material at high
magnetic fields.

This work is supported by the Department of Energy, the National
Science Foundation (NSF) and the State of Florida. JS acknowledges
EPSRC (UK) for support. We should
like to thank Albert Migliori and Steve Blundell for useful discussions.

\end{document}